\documentclass[10pt,journal]{IEEEtran} 
\usepackage{amssymb}
\usepackage{amsmath}
\usepackage{amsthm} % 如果要使用proof语句就必须要引入这个包

\usepackage{subfigure}  % 过时但还在用
\usepackage{graphicx}
\usepackage{flushend} % 自动平衡最后一页的双栏
\usepackage{algorithm}
\usepackage{algorithmic}

\begin{document}

\title{MD-AirComp+: Adaptive Quantization for Blind Massive Digital Over-the-Air Computation}

\author{Li Qiao,~\IEEEmembership{Member,~IEEE}, 
Yueqing Wang, 
Hanjun Jiang, 
Xinhua Liu, 
Yixuan Xing, 
\\
Yongpeng Wu,~\IEEEmembership{Senior Member,~IEEE}, 
and Zhen Gao,~\IEEEmembership{Senior Member,~IEEE}
\vspace{-12mm}

\thanks{L. Qiao is with the Department of Electrical and Electronic Engineering, The University of Hong Kong, Pokfulam Road, Hong Kong (e-mail: qiaoli@hku.hk).}

\thanks{Y. Wang and X. Liu are with the School of Information and Electronics, Beijing Institute of Technology, Beijing 100081, China.}

\thanks{H. Jiang, Y. Xing, and Z. Gao are with the School of Interdisciplinary Science, Beijing Institute of Technology, Beijing 100081, China (e-mail: gaozhen16@bit.edu.cn).}

\thanks{Y. Wu is with the Department of Electronic Engineering, Shanghai Jiao Tong University, Minhang, Shanghai 200240, China.}

% \thanks{Corresponding author: Zhen Gao.}

}
\maketitle

\begin{abstract}
Recent research has shown that unsourced massive access (UMA) is naturally well-suited for over-the-air computation (AirComp), as it does not require knowledge of each individual signal, as demonstrated by the massive digital AirComp (MD-AirComp) scheme proposed in \cite{ISIT,qiao2024massive}. The MD-AirComp scheme has proven effective in federated edge learning and is highly compatible with current digital wireless networks. However, it depends on channel pre-equalization, which may amplify computation errors in the presence of channel estimation inaccuracies, thus limiting its practical use. In this paper, we propose a blind {\it MD-AirComp+} scheme, which takes advantage of the channel hardening effect in massive multiple-input multiple-output (MIMO) systems. We provide an upper bound on the computation mean square error, analyze the trade-off between computation accuracy and communication overhead, and determine the optimal quantization level. Additionally, we introduce a deep unfolding algorithm to reduce the computational complexity of solving the underdetermined detection problem formulated as a least absolute shrinkage and selection operator optimization problem. Simulation results confirm the effectiveness of the proposed MD-AirComp+ framework, the optimal quantization selection strategy, and the low-complexity detection algorithm.
\end{abstract}\vspace{-2mm}

% \begin{document}
% \maketitle

\section{Introduction}
\label{sec1}
The integration of the Internet of Things (IoT) and Artificial Intelligence (AI), known as the Artificial Intelligence of Things (AIoT), represents a pivotal paradigm shift for next-generation wireless networks \cite{AIoT}. To realize ubiquitous intelligence, these networks are evolving into multi-dimensional integrated architectures—encompassing space, air, ground, and sea—to ensure seamless connectivity across diverse and complex environments \cite{liu2024near, liu2025toward}. At its core, the efficacy of AI, particularly deep learning (DL), relies on three essential pillars: data, computational power, and algorithms \cite{lecun2015deep}. AIoT ecosystems provide both abundant data from heterogeneous sensors and distributed computational resources, empowered by energy-efficient edge hardware that facilitates local processing \cite{wang2025empowering}. Consequently, a multitude of AIoT devices can execute distributed AI tasks, such as federated learning \cite{FL-IoT, AIoT-Mahdi, Mag-Deniz} and collaborative inference \cite{shlezinger2023collaborative, zhang2025distributed}. In these distributed settings, the communication paradigm is shifting from traditional bit-level transmission toward task-oriented and context-aware frameworks, where information exchange is increasingly dictated by the specific requirements of the underlying AI applications \cite{qiao2025token}. A fundamental challenge in this context is the efficient fusion of information from geographically dispersed processing units. This bottleneck lies at the intersection of massive-scale communication and computation, necessitating a seamless integration of both domains. To address this, over-the-air computation (AirComp) has emerged as a promising solution, consolidating communication and computation into a unified process to significantly enhance the overall performance of AIoT systems \cite{surveyAirComp}.

\subsection{Literature Review}
AirComp transforms the uplink multiple access channel into a sum computing mechanism. Depending on the coding and modulation techniques used, AirComp can be categorized into two main schemes: analog and digital.

\subsubsection{Analog AirComp}
In analog AirComp, each device's signal undergoes pre-equalization by applying the inverse of the uplink (UL) channel gain from the device to the base station (BS), which acts as the parameter server. The pre-equalized signals are then modulated onto the amplitude of the transmit waveform. Through simultaneous transmission from multiple devices, the receiver can directly obtain the sum of the local signals by exploiting the superposed waveform over the multiple access channel (MAC). These properties make AirComp well-suited for joint computation and communication across a large number of devices. As a result, analog AirComp has been widely adopted in federated learning and inference. 

For instance, in federated edge learning (FEEL), the local model updates from various devices are compressed and transmitted via AirComp, after which the BS optimizes the receiver by accounting for the computational results rather than the communication error rate \cite{Deniz2, G_Zhu, Deniz}. The use of multi-input multi-output (MIMO) systems for AirComp is explored in \cite{YuanmingMIMO}. The impact of imperfect device synchronization is investigated in \cite{Yulin}. The statistical properties of massive MIMO channels are leveraged to enable blind FEEL \cite{amiri2021blind, razavikia2024blind}. A comprehensive overview of FEEL can be found in \cite{ChenMZ}. {In addition to standard architectures, recent works have explored reconfigurable intelligent surfaces (RIS) to enhance AirComp in complex environments \cite{DAirComp}.} In the context of federated and collaborative inference tasks, local devices perform AI inference, and the inference results are aggregated over AirComp at the BS \cite{yilmaz2025private, zhou2025towards}. Additionally, the authors of \cite{lo2023collaborative, liu2023over} propose fusing intermediate feature elements via AirComp, which improves inference accuracy with limited communication overhead.

\subsubsection{Digital AirComp}
While analog AirComp shows significant potential, most current wireless networks, including those based on 3rd Generation Partnership Project (3GPP) standards, predominantly rely on digital communication protocols and hardware \cite{3GPP}. Consequently, both existing and future networks may not support the flexible modulation schemes required for analog AirComp. In digital communication systems, quantization—the process of transforming continuous values into discrete sets—represents an initial step. One-bit quantization is particularly well-suited for aggregation over the MAC, serving as the foundation for works such as \cite{Kaibin} and subsequent extensions to FEEL in \cite{FSKMV, DigFL, 1bitAttack}. These studies typically employ one-bit gradient quantization coupled with digital modulation—specifically BPSK or QPSK—at the edge devices, while the central server performs gradient decoding via majority voting. Furthermore, the authors of \cite{FSKMV} and \cite{DigFL} introduce non-coherent detection methods, enabling digital AirComp with multi-bit quantization without requiring channel state information (CSI). {Beyond communication efficiency, this paradigm has been further adapted to enhance the resilience of federated learning against Byzantine attacks \cite{1bitAttack}.} 

Despite its benefits, this approach necessitates multiple frequency allocations to represent different quantization levels, raising concerns about spectral efficiency. Moreover, although the convergence of FEEL with one-bit quantization is guaranteed by the framework of sign stochastic gradient descent (SignSGD), higher accuracy is essential for other tasks, such as federated inference \cite{zhou2025towards, lo2023collaborative}, multi-sensor localization \cite{ngo2024type}, and federated conformal prediction \cite{zhu2024federated}. Recent studies \cite{ChannelComp, SumComp, razavikia2024blind} have proposed coding schemes that facilitate function computation using superimposed digital constellations.

Furthermore, adaptive quantization plays a critical role in balancing the trade-offs between communication and computation \cite{li2017fundamental}, a concept also explored in various downstream tasks, such as federated learning \cite{AdapQuanFL,FedPAQ,FedDQ}. However, the design of adaptive quantization in digital AirComp remains underdeveloped.
{ Specifically, most existing schemes \cite{li2017fundamental, AdapQuanFL,FedPAQ,FedDQ} rely on the idealized assumption of error-free channels, which limits their applicability in practical wireless environments. Moreover, they are not designed to exploit the waveform superposition property of the MAC, making their extension to digital AirComp a non-trivial challenge.}

\subsubsection{Random Access and AirComp}
From a random access perspective, AirComp can be viewed as an approach that tightly integrates random access with computation-aware data transmission \cite{JSAC-Editor, gao2023compressive}. 
Such integration inherently relies on massive access technologies, which are designed to support massive connectivity under stringent latency and signaling constraints. 
In traditional sourced massive access, compressed sensing (CS)-based algorithms have been extensively developed for active device detection, channel estimation, and data detection across various scenarios \cite{ke2020compressive, qiao2021massive}.

{However, when the receiver focuses on the collective data rather than individual device identities, the paradigm shifts from sourced access to unsourced massive access (UMA) \cite{Yury,wu2020massive,li2022joint,ke2023next}. Moreover, UMA enables all devices to share a common random access codebook, offering both practicality and ease of implementation. A comprehensive introduction to UMA can be found in \cite{liva2024unsourced}. Since the receiver in UMA is only concerned with the transmitted messages rather than device identities, it naturally aligns with the objectives of computing tasks. Building on this synergy, the authors of \cite{ISIT, qiao2024massive} redesigned UMA for digital AirComp, resulting in the concept of massive digital AirComp (MD-AirComp), which was demonstrated to be effective both theoretically and through simulations for FEEL tasks.}

Interestingly, the concept of MD-AirComp is similar to type-based multiple access (TBMA) schemes introduced in \cite{mergen2006type}, although the random access codebook in \cite{mergen2006type} is orthogonal, leading to significantly higher communication overhead. More recently, the theoretical analysis of MD-AirComp and TBMA at a massive scale has been conducted in \cite{ngo2024type}. Moreover, the authors of \cite{zhu2024federated} applied TBMA in federated conformal prediction. Similar to most existing digital AirComp schemes, MD-AirComp \cite{ISIT, qiao2024massive} and the TBMA-based AirComp approaches in \cite{ngo2024type,zhu2024federated} employ channel pre-equalization and use fixed quantization levels, without accounting for variations in bandwidth and channel conditions. Such a design reduces flexibility and leads to degraded computation accuracy under constrained communication resources and imperfect CSI.

\subsection{Contributions}
To overcome the limitations of existing digital AirComp schemes, this work proposes the \emph{MD-AirComp+} framework, which offers the following key contributions.
\begin{itemize} 
\item{\bf Communication-efficient blind MD-AirComp method:}  
To eliminate the overhead and potential errors associated with channel pre-equalization, MD-AirComp+ exploits the statistical properties of massive MIMO. In this way, the receiver can mitigate the impact of unknown channels without requiring individual CSI from multiple devices. Moreover, the framework adaptively determines the optimal quantization level according to bandwidth and channel conditions, thereby achieving a balance between computing accuracy and communication overhead.
\item {\bf MSE analysis and optimal quantization level selection:}  
We analytically establish an upper bound on the computation mean squared error (MSE). Furthermore, we characterize the interplay among quantization level, bandwidth, and signal-to-noise ratio (SNR), and demonstrate the existence of an optimal quantization level that minimizes the computation MSE, thereby balancing computing precision with communication overhead.
\item {\bf Deep unfolding-aided low-complexity algorithm:}  
To address the high computational complexity of the receiver-side detection algorithm, we introduce a deep unfolding method, where a neural network is used to model key iterative parameters. This results in a 25-fold reduction in the number of iterations required to achieve convergence. 
\end{itemize}
\textit {Notation}: Boldface lower and upper-case symbols denote column vectors and matrices, respectively. For a matrix ${\bf A}$, ${\bf A}^T$, ${\bf A}^*$, ${\bf A}^H$, and ${\left\| {\bf{A}} \right\|_F}$ denote the transpose, complex conjugate, Hermitian transpose, and Frobenius norm of ${\bf{A}}$, respectively. For a vector ${\bf x}$, $\| {\bf x} \|_p$ denotes the ${\ell_p}$ norm of ${\bf x}$.  $[K]$ denotes the set $\{1,2,...,K\}$. ${\bf I}_K$ denotes the identity matrix with dimension $K\times K$.

\section{System Model}
\begin{figure}[t]
\centering
    \includegraphics[width=1\linewidth]{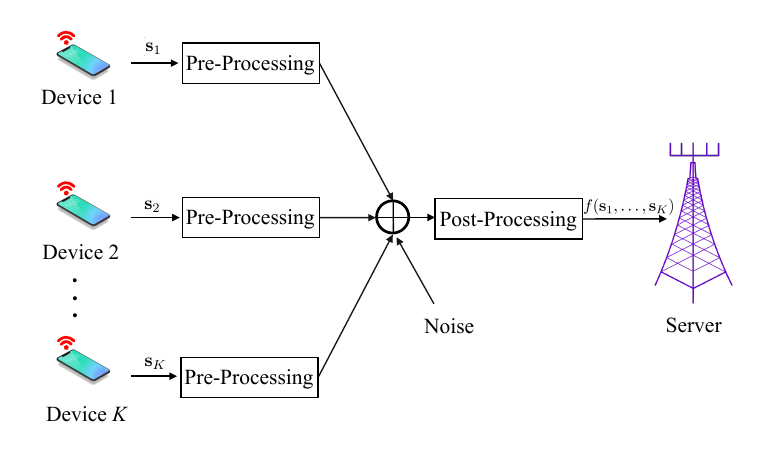}
\caption{Illustration of a typical digital AirComp scenario.}
\label{fig1}
\vspace{-3mm}
\end{figure}

As depicted in Figure \ref{fig1}, we consider a general digital AirComp framework in which $K$ distributed devices each hold local signals $\mathbf{s}_k$. These devices apply pre-processing operations, such as quantization and modulation, before transmitting the processed signals to a common edge server. The server’s objective is to compute a desired function, $f(\mathbf{s}_1, \ldots, \mathbf{s}_K)$, of all local observations—such as an average or weighted sum—by performing post-processing on the aggregated signals received over the multiple access channel. AirComp exploits the waveform superposition property of wireless channels to enable low-latency, bandwidth-efficient data aggregation in large-scale networks.

In addition to FEEL and federated inference, two other prominent application scenarios of digital AirComp are discussed below.

\noindent {\bf 1) Federated conformal prediction:} In federated conformal prediction, each client computes a local predictive probability vector for a given input sample \cite{humbert2023one}. Rather than transmitting raw data or full model parameters, clients transmit only these local prediction vectors; as shown in \cite{zhu2024federated}, they can be aggregated via digital AirComp to form calibrated prediction sets with guaranteed coverage. This approach enables reliable uncertainty quantification while substantially reducing communication overhead.

\noindent {\bf 2) Multi-sensor collaborative localization:}
In collaborative localization, multiple sensors (such as distributed access points, base stations, or vehicles) collect noisy measurements of a target's position \cite{ke2020massive}. Each sensor computes a local statistic (e.g., a likelihood score or coordinate estimate) and transmits it to the edge server for fusion \cite{qiao2023sensing}. By leveraging digital AirComp, the server can efficiently aggregate these statistics to derive an accurate global position estimate, as discussed in \cite{ngo2024type}. This approach enables scalable and timely localization in dense IoT networks.

\paragraph{Remark 1.}
While FEEL can tolerate some quantization errors during training due to the iterative nature of model updates \cite{G_Zhu,qiao2024massive}, inference-driven applications such as federated conformal prediction and multi-sensor localization demand higher reliability and accuracy, as highlighted in \cite{zhu2024federated, ngo2024type, yilmaz2025private}. Consequently, minimizing computational errors is generally the main optimization goal in digital AirComp schemes.

\subsection{MD-AirComp Overview}
By integrating communication and computation, MD-AirComp, as proposed in \cite{ISIT, qiao2024massive}, achieves low communication latency while remaining compatible with existing digital wireless networks. In this section, we first present the framework of MD-AirComp, and the enhanced \emph{MD-AirComp+} will be introduced in the subsequent section.

We consider a system with \(K\) devices, where the continuous source signal generated by the \(k\)-th device is denoted as \({\bf s}_{k} = [s_k^1, s_k^2,\dots,s_k^W]^T \in \mathbb{R}^{W}\), \(\forall k\in[K]\), with \(W\) representing the dimension of each source signal. Then, each source signal is stochastically quantized to \({\bf x}_{k}=[x_k^1, x_k^2,\dots,x_k^W]^T\in\mathbb{R}^{W}\) using a codebook \({\bf u}=[u_{1},\dots,u_{Q}]^T \in \mathbb{R}^{Q}\), where \(Q=2^{J}\) denotes the number of codewords. For each symbol \(i\in[W]\), the quantized signal is expressed as
\begin{align}\label{eq:quant}
x_{k}^i = \mathbf{u}^T{\bf z}_{k}^i,
\end{align}
where \({\bf z}_{k}^i \in \{0,1\}^{Q}\) is the one-hot index vector corresponding to the \(k\)-th source. Collecting all index vectors across \(W\) symbols yields \(\mathbf{Z}_k = [\mathbf{z}_k^{1}, \ldots, \mathbf{z}_k^{W}] \in \{0,1\}^{Q\times W}\). For notational simplicity, we focus on a single symbol in the subsequent analysis and omit the index \(i\).

During the uplink transmission, the received signal at the server $\mathbf{Y}\in\mathbb{C}^{L\times M}$ is given by
\begin{align}\label{eq:sysmod}
\mathbf{Y}=g\!\Big(\mathbf{P}\sum_{k=1}^{K}\mathbf{z}_k{\bf h}_k^T\Big)+\mathbf{N},
\end{align}
where \(\mathbf{P}\in\mathbb{C}^{L\times Q}\) denotes the random access preambles (or sensing matrix), \(L\) is the preamble length (i.e., the number of measurements), and \(\mathbf{N}\) is additive white Gaussian noise (AWGN) with variance \(\sigma^{2}\) per entry. According to \cite{qiao2024massive}, the function \(g(\cdot)\) implements pre-equalization at the transmitters to mitigate channel distortion $\mathbf{h}_k\in\mathbb{C}^{M}$, where $M$ denotes the number of antennas at the receiver. Based on \(\mathbf{Y}\), an estimate \(\widehat{\mathbf{z}}\) of the summation \(\sum_{k=1}^{K}\mathbf{z}_k\) is obtained, from which the estimated average signal is
\begin{align}\label{eq:summ}
\widehat{x}=\frac{1}{K}\mathbf{u}^T\widehat{{\bf z}}.
\end{align}
The true average signal is defined as \(\bar{s}\triangleq\frac{1}{K}\sum_{k=1}^{K}s_{k}\).

\paragraph{Stochastic Uniform Quantizer.}
We employ the widely adopted stochastic uniform quantizer \(\mathcal{Q}_{\bf u}(\cdot)\) based on the predefined codebook \({\bf u}=[u_{1},\dots,u_{Q}]^T\) with \(Q=2^{J}\) quantization levels. For each entry of a \(W\)-dimensional source vector \({\bf s}_{k}=[s_{k}^{1},\dots,s_{k}^{W}]^T\), the quantizer randomly assigns the value to one of two adjacent codewords according to its relative position between them. Specifically, for the \(i\)-th entry \(s_{k}^{i}\), $i\in[W]$, we first determine an index \(l\in\{1,\dots,Q-1\}\) such that
\[
u_{l}\leq s_{k}^{i}<u_{l+1}.
\]
Then the stochastic quantization output \(x_{k}^{i}\) is given by
\begin{align}\label{eq:quant2}
x_{k}^{i}=
\begin{cases}
u_{l+1}, & \text{with probability } 
\displaystyle\frac{s_{k}^{i}-u_{l}}{u_{l+1}-u_{l}},\\[1ex]
u_{l}, & \text{otherwise}.
\end{cases}
\end{align}
Equivalently, the index vector \(\mathbf{z}_{k}^{i}\in\{0,1\}^{Q}\) is a one-hot vector with the position corresponding to the randomly chosen codeword. For inputs that are outside the codebook range, we define \(\mathcal{Q}_{\bf u}(s_{k}^{i})\) as the nearest boundary codeword, i.e., the codeword corresponding to the extreme values of the codebook. This stochastic quantization ensures that the expectation of the quantized value equals the original continuous value within each quantization interval \cite{suresh2017distributed}.

It is worth noting that MD-AirComp can also leverage vector quantization (VQ) to further improve communication efficiency, as discussed in \cite{ISIT, qiao2024massive}. In this paper, however, we mainly focus on scalar quantization, while the extension to VQ is demonstrated in the simulation part.

\begin{figure}[t]
\centering
    \includegraphics[width=1\linewidth]{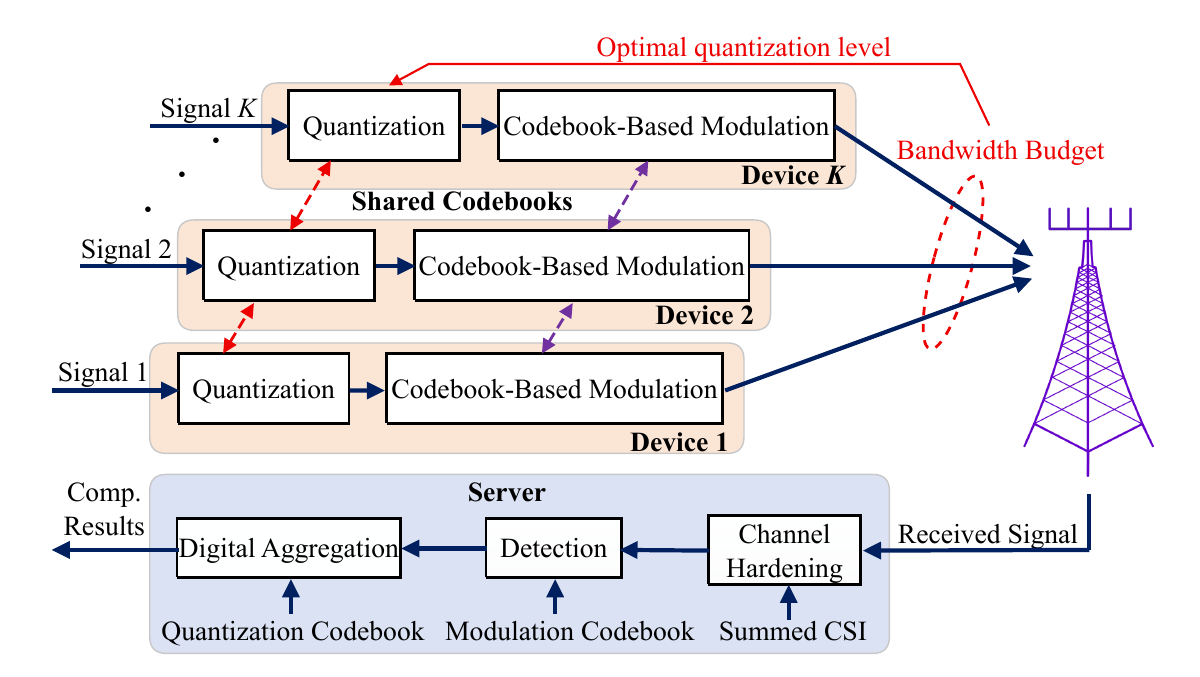}
\caption{Illustration of the proposed MD-AirComp+ scheme.}
\label{fig2}
\vspace{-5mm}
\end{figure}

\section{Proposed MD-AirComp+ Scheme}

This subsection highlights two critical limitations of the original MD-AirComp scheme and motivates the enhanced \emph{MD-AirComp+} design.

\textbf{1) Sensitivity to channel reciprocity and estimation errors.} 
The original MD-AirComp assumes reciprocity between the uplink and downlink channels. 
Thus, the estimated downlink channel is used at the transmitter for pre-equalization. 
However, as the channel estimation error increases, i.e., the function $g(\cdot)$ in (\ref{eq:sysmod}) cannot perfectly pre-equalize the channel, the accuracy of the aggregated summation inevitably degrades. 
As for inference tasks, e.g., \cite{zhu2024federated, ngo2024type,yilmaz2025private}, such estimation errors directly impair performance and become unacceptable.

\textbf{2) Trade-off between quantization and computing accuracy.} 
General downstream tasks such as inference typically demand much higher quantization fidelity than FEEL: finer quantization improves accuracy but also increases the dimensionality of the target signal \(\mathbf{z}\). 
When combined with limited system resources (e.g., the preamble length \(L\)), this higher signal dimension leads to more detection errors, which in turn exacerbate reconstruction errors. 
Therefore, the trade-off between quantization precision and computational accuracy poses a critical challenge for the original MD-AirComp scheme when applied to general computing tasks.

Motivated by these challenges, we propose the \emph{MD-AirComp+} scheme, which is specifically designed to tackle the lack of accurate individual CSI for different users, and meet the quantization fidelity requirements for general computing tasks, while mitigating the impact of higher signal dimensions.

\begin{figure}[t]
\centering
    \includegraphics[width=1\linewidth]{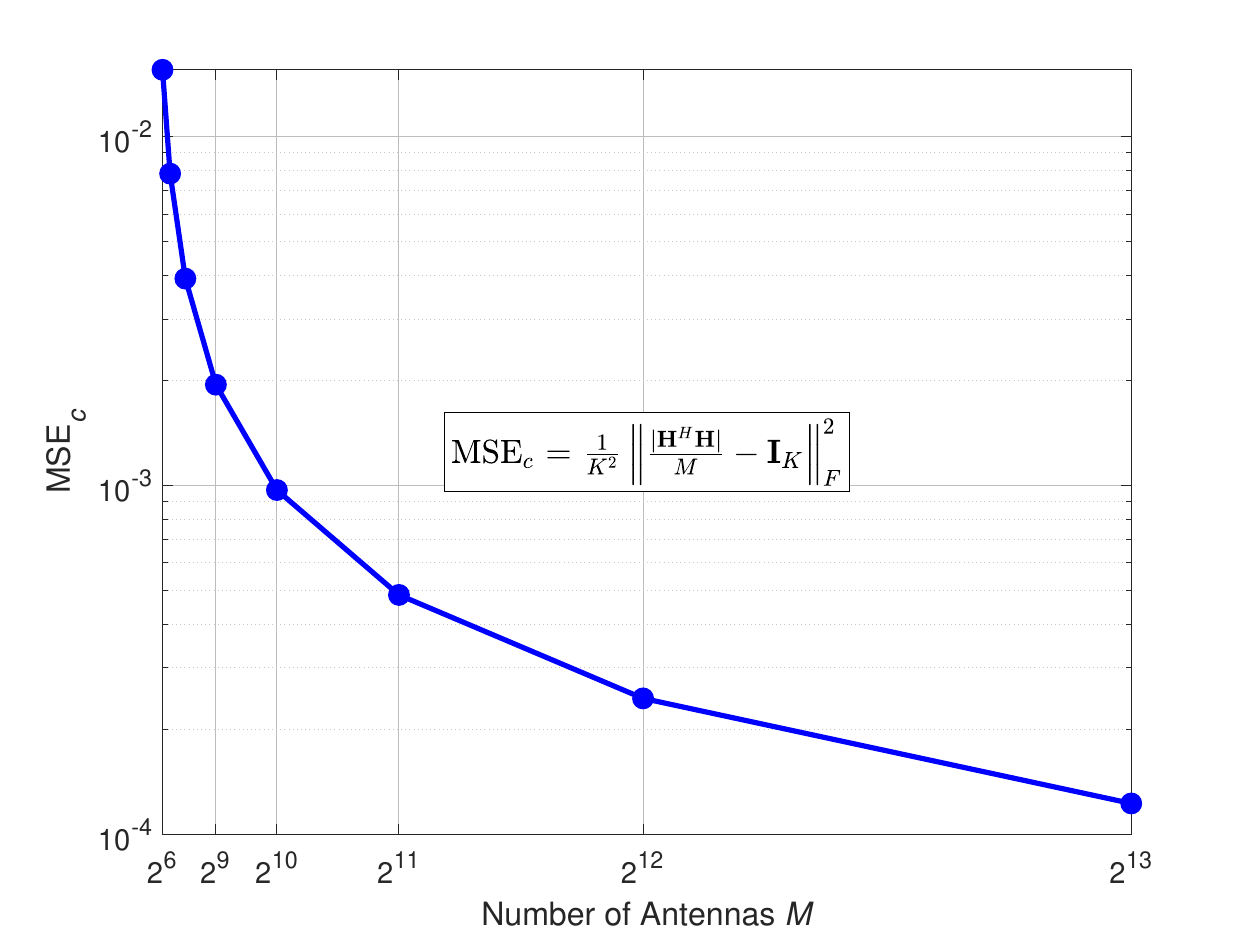}
\caption{Illustration of the channel hardening effect in massive MIMO system.}
\label{fig3}
\vspace{-5mm}
\end{figure}

\subsection{Blind Channel Mitigation}
In our system, we leverage the principle of channel hardening, similar to the approach used in \cite{amiri2021blind, razavikia2024blind}. In a massive MIMO system, as the number of antennas at the BS increases, the channel fading becomes more predictable and stable, leading to channel hardening \cite{lu2014overview}. This phenomenon causes the inner products of the channel vectors for the same user to approach a large constant value (related to \(M\)), while the inner products between different users' channel vectors approach zero, implying \textit{orthogonality} between them. 

In our approach, we assume that at the initial time slot, \(K\) devices transmit a common pilot signal. The server then estimates the summation of the \(K\) channel vectors, i.e., \(\overline{\bf h}= \sum_{k=1}^{K} {\bf h}_k\), where each \({\bf h}_k\) follows a complex Gaussian distribution (Rayleigh channel assumption). {Estimating $\bar{\mathbf{h}}$ via a common pilot requires much lower overhead than individual CSI.} Without loss of generality, we assume that \(\overline{\bf h}\) is perfectly estimated. By multiplying \(\overline{\bf h}^*\) to the system model in (\ref{eq:sysmod}), we obtain
\begin{align}\label{eq:sysmod2}
\frac{\mathbf{Y}\overline{\bf h}^*}{M} &= \mathbf{P} \sum_{k=1}^{K} \mathbf{z}_k \frac{{\bf h}_k^T \overline{\bf h}^*}{M} + \frac{\mathbf{N} \overline{\bf h}^*}{M}\\ \nonumber
& \overset{(a)}{\approx} \mathbf{P} \sum_{k=1}^{K} \mathbf{z}_k + \frac{\mathbf{N} \overline{\bf h}^*}{M},
\end{align}
where \( \overset{(a)}{\approx} \) is valid when \( M \) is sufficiently large. As illustrated in Figure \ref{fig3}, the y-axis represents the channel hardening metric $\text{MSE}_c = \frac{1}{K^2} \left\| \frac{\mathbf{H}^H \mathbf{H}}{M} - \mathbf{I}_K \right\|_F^2$, where $K = 10$ and $\mathbf{H} = [\mathbf{h}_1, \mathbf{h}_2, \dots, \mathbf{h}_K] \in \mathbb{C}^{M \times K}$. It is evident from Figure \ref{fig3} that $\text{MSE}_c$ decreases monotonically as the number of antennas increases, thereby empirically validating the channel hardening effect. {In practical deployments, such as 5G NR systems with 256 antennas, this effect is sufficiently pronounced to stabilize the composite channel gain. The impact of non-ideal channel hardening under finite antenna counts will be evaluated in detail in the simulation section.}

\subsection{LASSO-Based Sparse Signal Detection}
For notational simplicity, we reformulate (\ref{eq:sysmod2}) as
\begin{align}
\overline{\mathbf{y}} &= \mathbf{Pz} + \overline{\mathbf{n}},
\end{align}
where \( \overline{\mathbf{y}} = \frac{\mathbf{Y}\overline{\mathbf{h}}^*}{M} \) and \( \overline{\mathbf{n}} = \frac{\mathbf{N} \overline{\mathbf{h}}^*}{M} \). This equation represents a sparse signal detection problem, where \( \overline{\mathbf{y}} \) is the observed noisy measurement, \( \mathbf{Pz} \) is the signal model, and \( \overline{\mathbf{n}} \) is the noise.

To solve this problem, we employ the iterative shrinkage-thresholding algorithm (ISTA), which is commonly used to solve the least absolute shrinkage and selection operator (LASSO) problem. The LASSO problem is formulated as:
\begin{align}\label{eq:opt}
\hat{\bf z} &= \arg \min_{\bf z} \frac{1}{2} \| \overline{\mathbf{y}} - \mathbf{Pz} \|^2 + \rho \| {\bf z} \|_1,
\end{align}
where \( \rho \) is a regularization parameter and \( \| {\bf z} \|_1 \) is the \( \ell_1 \)-norm that enforces sparsity on the solution vector \( {\bf z} \).

We begin by recalling the formulation of ISTA \cite{gregor2010learning}. The original algorithm aims at solving the LASSO problem through the iterative update equations:
\begin{align}\label{eq:ISTA_ite}
\mathbf{z}^{(t+1)} &\leftarrow T_{\beta} \left( \mathbf{z}^{(t)} - \mu \mathbf{P}^H (\mathbf{P}\mathbf{z}^{(t)} - \overline{\mathbf{y}}) \right),
\end{align}
where \( T_{\beta} \) is the soft-thresholding operator defined as:
\begin{align}
T_{\beta}(x) &= \text{sign}(x) \cdot \max(0, |x| - \beta).
\end{align}
The parameter \( \mu \) represents the step size. The convergence of ISTA can be analyzed when the step size \( \mu \) satisfies the condition:
\begin{align}
\mu &\leq \frac{1}{\lambda_{\text{max}}(\mathbf{P}^H \mathbf{P})},
\end{align}
where \( \lambda_{\text{max}}(\mathbf{P}^H \mathbf{P}) \) is the largest eigenvalue of \( \mathbf{P}^H \mathbf{P} \). This condition ensures the stability and convergence of the algorithm. Thus, we set \( \mu = \frac{1}{\lambda_{\text{max}}(\mathbf{P}^H \mathbf{P})} \).

The iterative process is repeated until convergence, providing an estimate for the sparse vector \( \hat{\mathbf{z}} \), which corresponds to the sparse signal we are detecting.

\section{Communication and Accuracy Trade-off Analysis}

We now decompose the recovery error between the original signal average \(\bar s\) and the reconstructed average \(\widehat{x}\) into two terms.

First, insert and subtract the quantized average \(\bar x \triangleq \frac{1}{K}\sum_{k=1}^K x_k = \frac{1}{K}\mathbf{u}^T\sum_{k=1}^K \mathbf{z}_k\):
\begin{align}
  \bar s - \widehat{x}
  &= (\bar s - \bar x) + (\bar x - \widehat{x}) \nonumber\\
  &= \underbrace{\frac{1}{K}\sum_{k=1}^K (s_k - x_k)}_{\text{(I): quantization term}}
  \;+\;
  \underbrace{\mathbf{u}^T\left(\frac{1}{K}\sum_{k=1}^K \mathbf{z}_k -  \frac{1}{K}\widehat {\mathbf{z}}\right)}_{\text{(II): detection term}}.\label{eq:two_terms}
\end{align}

Now, we can compute the total error by taking the squared norm and expectation
\begin{align}\label{eq:mse}
    \text{MSE} &\triangleq \mathbb{E}\|\bar s - \widehat{x}\|_2^2 \\ \nonumber
  &= \mathbb{E}\Big\| \underbrace{\frac{1}{K}\sum_{k=1}^K (s_k - x_k)}_{a}
        \;+\;
        \underbrace{\mathbf{u}^T\frac{1}{K} (\sum_{k=1}^K\mathbf{z}_k - \widehat {\bf z})}_{b}\Big\|_2^2.
\end{align}
  
Expanding the squared norm, we get the following additive upper bound:
\begin{equation}\label{eq:E_split}
  \text{MSE} \le 2\,\mathbb{E}\|a\|_2^2 + 2\,\mathbb{E}\|b\|_2^2.
\end{equation}
In the following, we analyze the two error terms one-by-one.

\subsection{Step 1 — quantization error}
The scalar quantization step size is \(\Delta = \frac{2R}{2^J}\),
if each scalar of $s_k$ lies in $[-R,R]$. The per-scalar mean squared quantization error of a uniform quantizer is
\begin{align} \label{eq:quant_bound}
      \mathbb{E}\big[\|s_{k}-x_{k}\|_2^2\big] = \frac{\Delta^2}{12} = \frac{(2R)^2}{12\cdot 4^J} = \frac{R^2}{3\cdot 4^J}\triangleq \mathcal{Q}(J),
\end{align}
where we explicitly emphasize the dependence on $J$. 

Due to the linearity and identical distribution of quantization errors across sources, we can derive
\begin{align} \label{eq:a_upper_bound}
      \mathbb{E}\|a\|_2^2
  &= \mathbb{E}\left\|\frac{1}{K}\sum_{k=1}^K (s_k - x_k)\right\|_2^2\\ \nonumber
  &= \frac{1}{K^2}\sum_{k=1}^K \mathbb{E}\|s_k - x_k\|_2^2
  = \frac{1}{K}\,\mathcal{Q}(J).
\end{align}

\subsection{Step 2 — LASSO for recovering $\mathbf{z}$ and its error bound}

Under the typical assumptions, e.g., $\mathbf{P}$ satisfies restricted isometry property (RIP) conditions and noise is Gaussian, we can invoke a standard LASSO-type error bound. { In our model, each device maps its signal to a single codeword; thus, the number of non-zero elements in the frequency vector $\mathbf{z}$ is $\Vert \mathbf{z} \Vert_0 \le \min(K, 2^J) \le K$. While the actual sparsity $\Vert \mathbf{z} \Vert_0$ may be smaller than $K$ due to ``collisions'' (multiple devices falling into the same quantization bin) at low resolutions, we use $K$ as a distribution-agnostic upper bound to derive a robust performance guarantee.} Hence, a representative bound
\begin{equation}\label{eq:lasso_matrix_bound}
  \mathbb{E}\big\|\widehat{\mathbf{z}} - \mathbf{z}\big\|_2^2
  \le
  C_0\frac{\, K\, \sigma^2 \,2^J\log 2^J}{L},
\end{equation}
holds with high probability provided that $L = \Omega(K \log(2^J/K))$, where $C_0$ is a constant depending on the estimator and sensing matrix properties \cite{candes2013well}.

Then, according to the Cauchy–Schwarz inequality, its expected squared norm can be bounded as
\begin{align} \label{eq:b_upper_bound}
    \mathbb{E}\|b\|_2^2 
  &\le \|\frac{1}{K}\mathbf{u}^T\|_2^2 \,\,\mathbb{E}\| \widehat{\mathbf{z}}- \mathbf{z}\|_2^2.\nonumber \\
&\le C_0\|\mathbf{u}\|_2^2\frac{ \sigma^2\,2^J\log 2^J}{KL}.
\end{align}

\paragraph{Total bound.}
By applying (\ref{eq:a_upper_bound}) and (\ref{eq:b_upper_bound}) into (\ref{eq:E_split}), we can obtain the total upper bound as
\begin{align}\label{eq:E_total}
  \text{MSE} &\lesssim \underbrace{\frac{2}{K}\,\mathcal{Q}(J)}_{\text{quantization}} \;+\;
\underbrace{C_0\|\mathbf{u}\|_2^2\frac{ \,2\sigma^22^J\log 2^J}{KL}}_{\text{detection}}\\ \nonumber
  &=\frac{2R^2}{3K4^J} + C_0\|\mathbf{u}\|_2^2\frac{ \,2\sigma^22^J\log 2^J}{KL}.
\end{align}

\begin{algorithm}[!t]
\small
\caption{Proposed MD-AirComp+ Scheme}
\label{alg1}
\begin{algorithmic}[1]
\REQUIRE Preamble length $L$ (determined by bandwidth/latency constraints); Noise variance $\sigma^2$; Number of devices $K$.
\ENSURE Aggregated result at the server.
\STATE \textbf{Initialization:} The server initiates a network-wide beacon transmission for time and frequency synchronization.
\STATE \textbf{Channel Estimation:} All devices transmit a common pilot signal during the initial slot. The server estimates the composite channel via least-squares (LS) or linear minimum mean square error (LMMSE) based on the received pilots.
\STATE \label{line:quant} \textbf{Adaptive Quantization:} Given $L$, $K$, $\sigma^2$, and the detection algorithm, determine the optimal quantization level $Q$ according to (\ref{eq:bound_f}).
\STATE \textbf{Codeword Selection:} Each device selects modulation codewords ${\bf P}{\bf z}_k$ based on the quantization in Step~\ref{line:quant}.
\STATE \textbf{Simultaneous Transmission:} All active devices transmit their data-embedded preambles simultaneously over the uplink; the server receives the superposed signals.
\STATE \label{line:postproc}{\textbf{Post-processing:} The server multiplies $\mathbf{y}$ with $\bar{\mathbf{h}}^H/M$ to exploit the channel hardening effect.}
\STATE {\textbf{Aggregation:} The server solves the sparse recovery problem, e.g., via (\ref{eq:opt}), to find $\hat{\mathbf{z}}$, then computes the final estimated average $\hat{x} = \frac{1}{K}\mathbf{u}^T\hat{\mathbf{z}}$.}

\end{algorithmic}
\end{algorithm}

\subsection{Optimal $Q$ Analysis under Preamble Length $L$ and SNR Constraints}

Since the number of quantization levels is $Q = 2^J$, the MSE upper bound can be approximated as
\begin{align}\label{eq:bound_f}
    \text{MSE} \lesssim \frac{2R^2}{3KQ^2} + C_0 \|\mathbf{u}\|_2^2 \frac{2\sigma^2Q \log Q \, }{K L}.
\end{align}
{It is evident that the MSE diverges as $Q \to 0^+$ or $Q \to +\infty$, implying the existence of an optimal quantization level $Q^*$ that minimizes the total error. Since the second derivative of the MSE with respect to $Q$ is positive for all $Q>0$, the objective function is strictly convex, ensuring that any critical point corresponds to a unique global minimum. 

The MSE performance is governed by a fundamental trade-off between quantization resolution and detection robustness under a fixed preamble length $L$. 
Specifically, as $Q$ increases, the quantization distortion decreases due to the finer resolution fidelity. 
However, since $L$ represents a finite bandwidth budget, a larger $Q$ renders the sparse recovery problem increasingly underdetermined as the sensing matrix $\mathbf{P} \in \mathbb{C}^{L \times Q}$ becomes ``fatter". 
This expansion of the hypothesis space causes the detection error to grow, eventually outweighing the gains in quantization precision. 
Although our analytical bound (\ref{eq:lasso_matrix_bound}) uses $K$ as a conservative proxy for the sparsity $\|\mathbf{z}\|_0 \le K$, it accurately captures the U-shaped scaling law and the existence of an optimal $Q^*$. 
The optimal level thus represents the equilibrium point where the marginal improvement in quantization accuracy is perfectly balanced against the escalating risk of detection failure under limited observation resources.
}

Finally, the proposed \textit{MD-AirComp+} scheme is summarized in Algorithm \ref{alg1}. The key difference compared to MD-AirComp is the handling of CSI and the quantization parameter selection according to the given bandwidth and the noise level.

\section{Low-Complexity Detection Algorithm Design}
While the ISTA provides a principled approach to solving the LASSO problem, its convergence rate can be relatively slow because both the step size $\mu$ and the thresholding parameter $\beta$ are fixed during iterations. Inspired by the idea of algorithm unfolding, the learned ISTA (LISTA) \cite{gregor2010learning} accelerates the convergence by treating the parameters and operators involved in the ISTA update as learnable. Specifically, LISTA learns the optimal step sizes and thresholding parameters at each iteration from data, and generalizes the fixed matrices in ISTA into trainable linear transforms. This data-driven adaptation allows LISTA to achieve faster convergence and improved recovery accuracy compared to the traditional ISTA \cite{shlezinger2023model}.

We adopt the unfolded form of ISTA by parameterizing the iteration-dependent step sizes, thresholds, and linear operators. Given the ISTA formulation (\ref{eq:ISTA_ite}), LISTA replaces the fixed parameters with learnable ones
\begin{align}\label{eq:LISTA_ite}
\mathbf{z}^{(t+1)} &\leftarrow 
T_{\beta^{(t)}}\!\left( 
\mathbf{z}^{(t)} - \mu^{(t)}\!\left(\mathbf{B}\mathbf{z}^{(t)} + \mathbf{A}\overline{\mathbf{y}}\right) 
\right),
\end{align}
where $\mathbf{A}$, $\mathbf{B}$, $\{\beta^{(t)},\mu^{(t)}\}$ are all trainable parameters learned from data. The soft-thresholding operator $T_{\beta^{(t)}}(\cdot)$ is applied element-wise at each iteration. The adopted ISTA and LISTA algorithms are summarized in Algorithm \ref{alg2}.
{Unlike ISTA which uses a conservative fixed step size to guarantee convergence for any sparse signal, LISTA learns the underlying distribution of the sensing matrix $\mathbf{P}$ and the data. By employing data-driven weights, it can effectively ``navigate'' the optimization landscape more efficiently, resulting in the 25-fold complexity reduction observed in Figure 4.}

\paragraph{Remark 2.}
While this work does not aim to optimize the detection module itself, it is worth noting that advanced message-passing and Bayesian inference frameworks exploiting prior statistical information, such as \cite{qiao2022joint}, are in principle capable of handling the detection challenges encountered in blind MD-AirComp systems. Integrating such approaches with the proposed quantization-adaptive framework is an interesting direction for future work.

\begin{algorithm}[!t]
\small
\caption{ISTA and LISTA Update Rules}
\label{alg:ISTA_LISTA}
\begin{algorithmic}[1]  % [1]表示自动编号
\REQUIRE Measurement $\overline{\mathbf{y}}$, sensing matrix $\mathbf{P}$, maximum iteration $T$ and $T'$ for ISTA and LISTA.
\ENSURE Estimated sparse signal $\mathbf{z}^{(T)}$
\vspace{0.5em}
\STATE \textbf{ISTA:}
\STATE Initialize $\mathbf{z}^{(0)}=\mathbf{0}$; set step size $\mu=\frac{1}{L}$ with $L=\lambda_{\text{max}}(\mathbf{P}^H\mathbf{P})$;
\FOR{$t=0$ \TO $T-1$}
    \STATE $\mathbf{z}^{(t+1)} \leftarrow T_{\beta}\!\left(\mathbf{z}^{(t)} - \mu \mathbf{P}^H\big(\mathbf{P}\mathbf{z}^{(t)}-\overline{\mathbf{y}}\big)\right)$;
\ENDFOR
\vspace{0.5em}
\STATE \textbf{LISTA:}
\STATE Initialize $\mathbf{z}^{(0)}=\mathbf{0}$; learnable $\{\mathbf{A},\mathbf{B},\beta^{(t)},\mu^{(t)}\}$ from training data;
\FOR{$t=0$ \TO $T-1$}
    \STATE $\mathbf{z}^{(t+1)} \leftarrow 
    T_{\beta^{(t)}}\!\left(
    \mathbf{z}^{(t)} - \mu^{(t)}\big(\mathbf{B}\mathbf{z}^{(t)}+\mathbf{A}\overline{\mathbf{y}}\big)
    \right)$;
\ENDFOR
\STATE \textbf{return} $\mathbf{z}^{(T)}$.
\end{algorithmic}
\label{alg2}
\end{algorithm}

\section{Simulation Results}
\label{sec:simulation}
In the simulations, the number of active devices is set to \(K = 10\), where each device generates a random value drawn from a uniform distribution over the interval \([0,1]\). The choice of \([0,1]\) is general: it can represent probabilistic outcomes of inference tasks such as classification \cite{zhu2024federated}, or alternatively, coordination within a normalized square area \cite{ngo2024type}. The modulation codebook \( \mathbf{P_0} \) is a fixed matrix, where each element is drawn from a complex Gaussian distribution, with dimensions \( 60 \times 256 \). This codebook is shared between the transmitter and the receiver. A subset of the codebook, denoted as \( \mathbf{P} \), is selected from \( \mathbf{P}_0 \), corresponding to a subarray of size \( L \times Q \), where the parameters \( L \) and \( Q \) are adjustable. The number of antennas is set to \( M = 1024 \), unless otherwise specified. The noise is modeled as complex Gaussian noise, and the massive MIMO channel undergoes Rayleigh fading. We consider the MSE metric as defined in (\ref{eq:mse}).

\begin{figure}[t]
\vspace{-2mm}
\centering
    \includegraphics[width=1\linewidth]{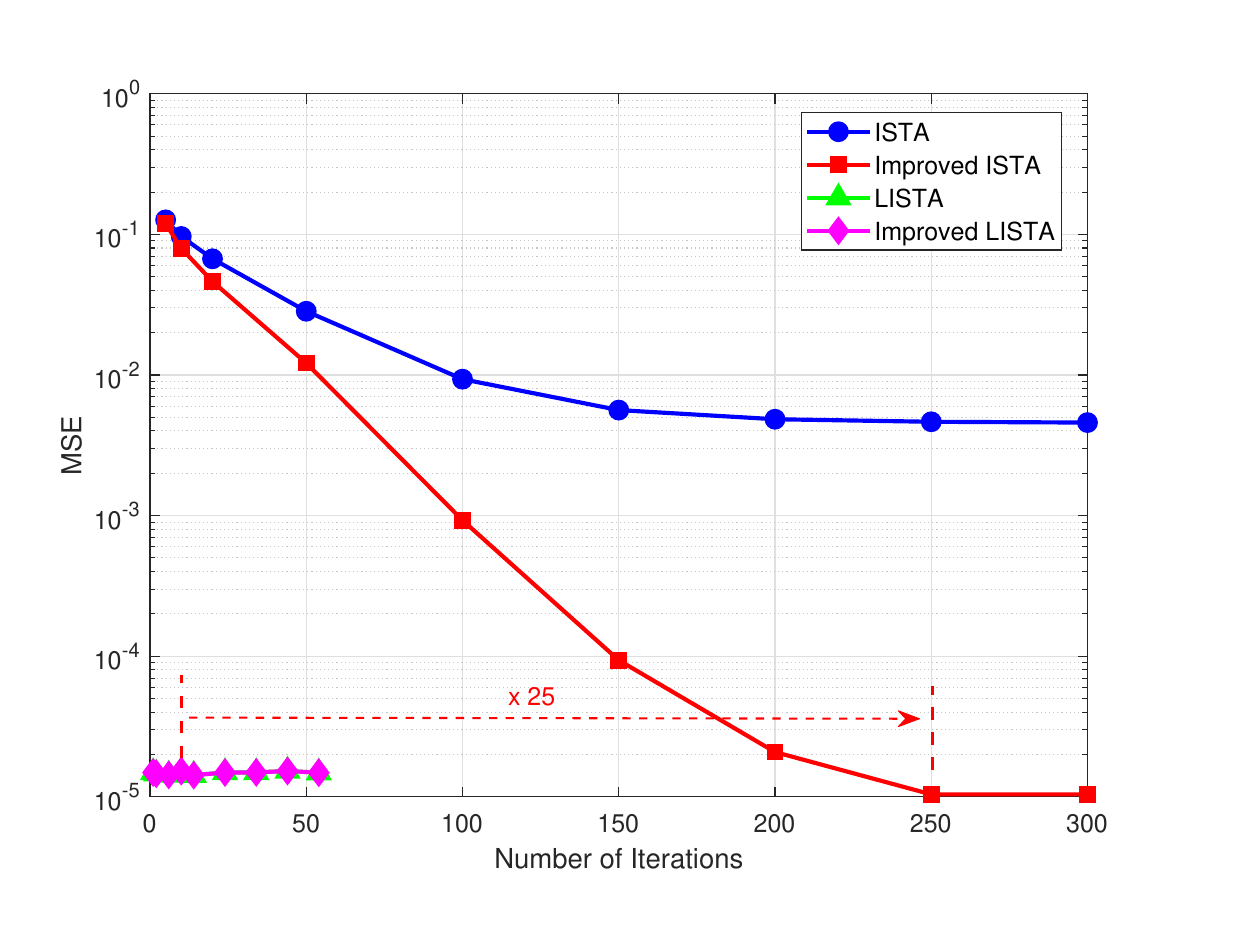}
\caption{Convergence and MSE comparison: ISTA versus LISTA and their improved variants}
\label{fig:convg}
\vspace{-5mm}
\end{figure}

\subsection{Convergence of the LISTA Algorithm}
Figure \ref{fig:convg} illustrates the convergence behavior and MSE performance of ISTA, LISTA, and their improved variants, namely ``Improved ISTA'' and ``Improved LISTA.'' In this case, we set \( L = 25 \), \( Q = 32 \), and \( \text{SNR} = 10 \, \text{dB} \). The improved versions incorporate prior information to enhance performance further. Specifically, because the non-zero elements in \( \mathbf{z} \) are integers, we apply rounding to the estimated values of \( \widehat{\mathbf{z}} \) after ISTA/LISTA estimation. It is evident from the results that LISTA converges after just 10 iterations, whereas ISTA requires 25 times more iterations. Furthermore, LISTA naturally incorporates the integer constraint through parameter training, while ISTA fails to do so, leading to a higher MSE plateau. After applying the improved ISTA, the final convergence plateau slightly exceeds that of LISTA, but at a significantly higher computational cost. Therefore, LISTA proves to be an effective method in applications where lower computational overhead is desired, achieving over 25 times reduction in complexity with only a minor loss in detection accuracy. In the subsequent analysis, we use the ``Improved ISTA'' algorithm with 300 iterations, which ensures optimal performance.

\begin{figure}[t]
    \centering
    % --- Left Subfigure ---
    \subfigure[]{
    \includegraphics[width=1\linewidth]{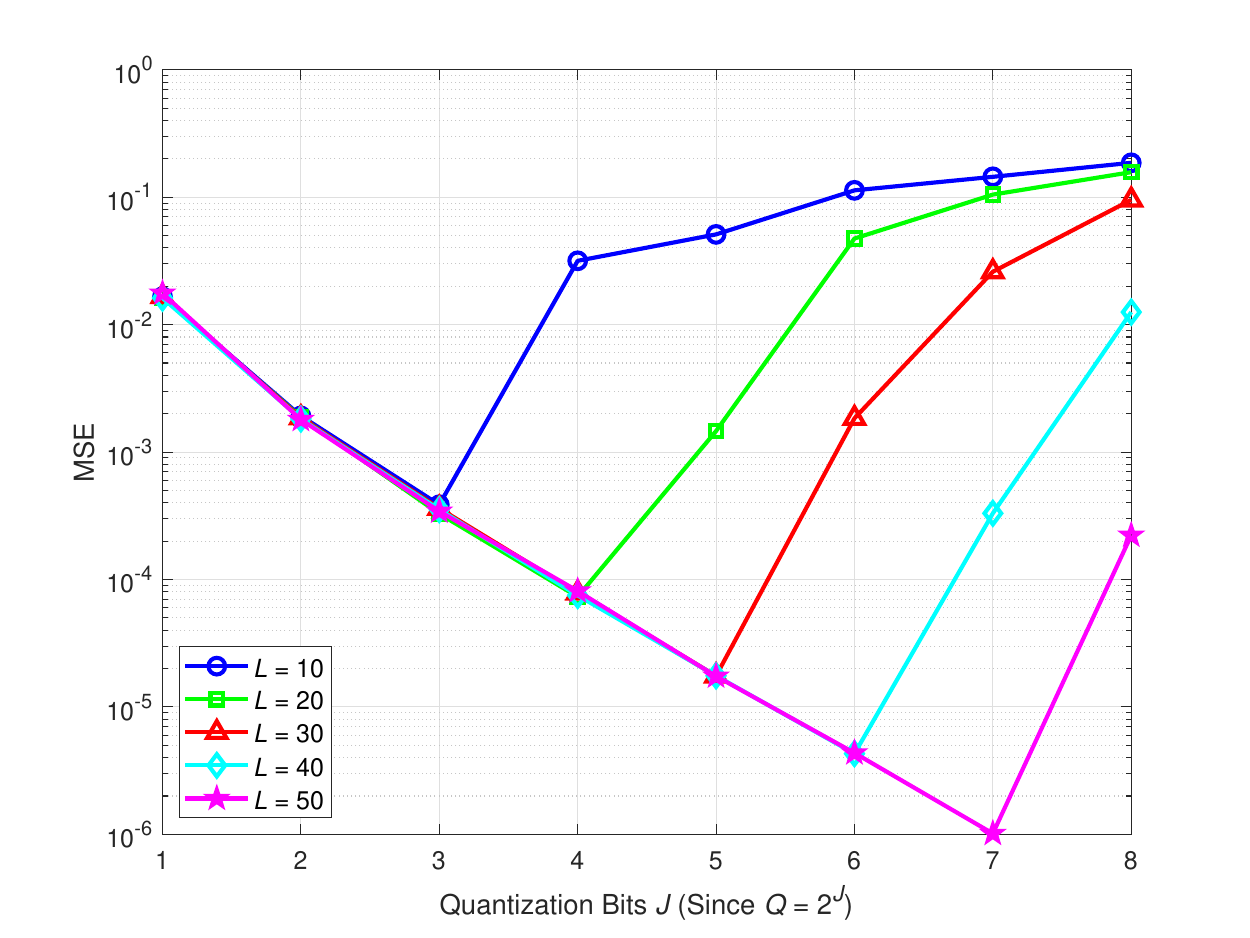}
        \label{fig:optQ1}
    }
    % --- Right Subfigure ---
    \subfigure[]{
    \includegraphics[width=1\linewidth]{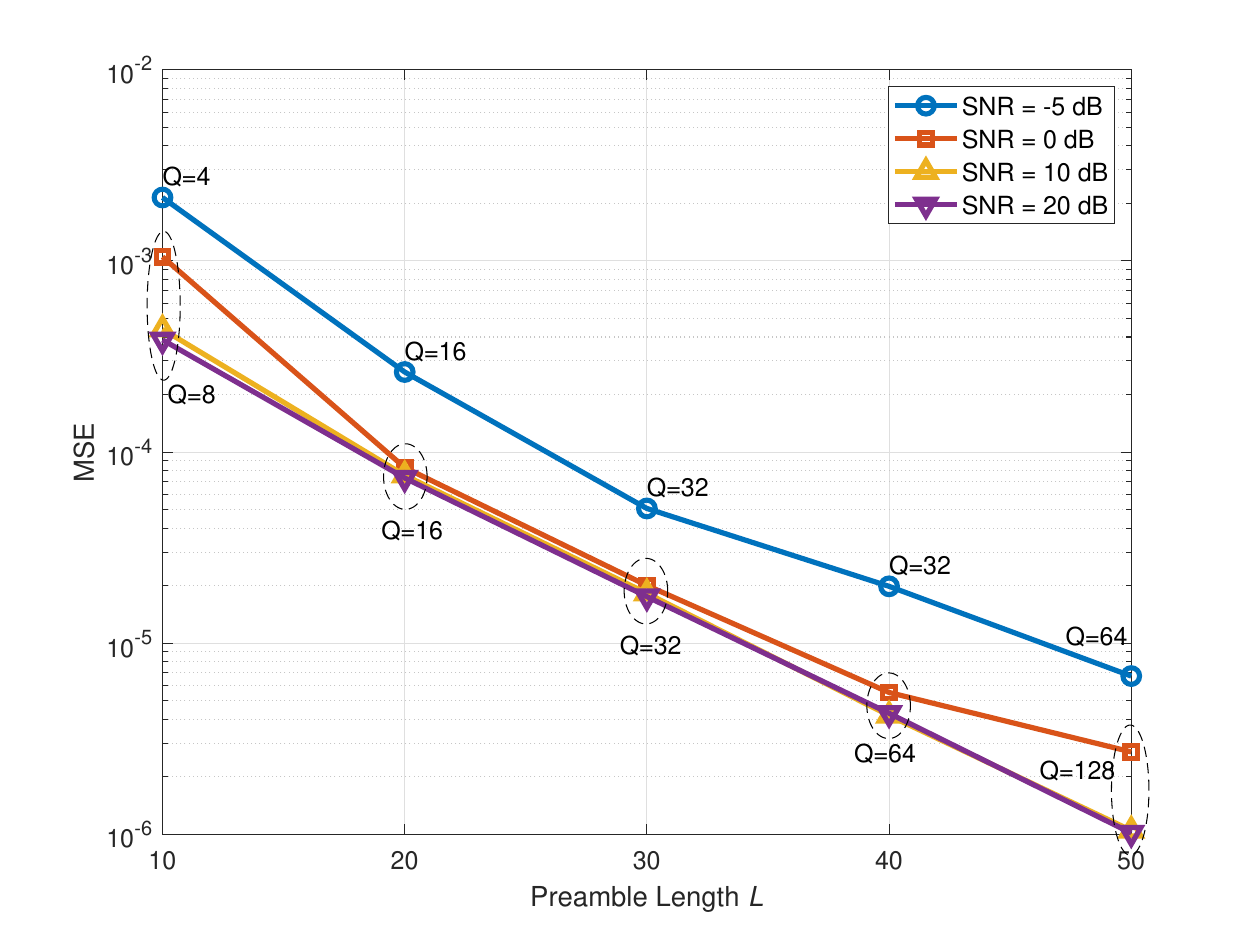}
        \label{fig:optQ2}
    }
    \vspace{-2mm}
    \caption{(a) MSE versus the number of quantization levels \( Q \), with fixed \( L \) and SNR = 20 dB. (b) Optimal number of quantization levels \( Q^* \) versus preamble length \( L \).}
    \label{fig:optQ}
    \vspace{-5mm}
\end{figure}

\subsection{Optimal \( Q \) Selection}
{ Figure \ref{fig:optQ} illustrates the impact of quantization levels $Q \in \{2^1, \dots, 2^8\}$, preamble length $L$, and SNR on the MSE performance. 
As shown in Figure \ref{fig:optQ1}, with SNR set to $20$ dB, the MSE exhibits a prominent U-shaped trend as $Q$ increases. 
This reveals a fundamental design trade-off: the total error is governed by a quantization-limited region on the left, where a coarse codebook causes high distortion, and a detection-limited region on the right, where a larger $Q$ increases the dimensionality of the sparse vector $\mathbf{z}$, thereby exacerbating recovery errors under limited preamble resources. 
The optimal $Q^*$ thus represents the equilibrium point that minimizes the total MSE for a given $L$. 
Specifically, Figure \ref{fig:optQ2} confirms that as the bandwidth (preamble length $L$) increases, the optimal MSE decreases significantly, highlighting the practical benefits of our adaptive quantization strategy. 
Notably, the preamble length $L$ exerts a more dominant influence on $Q^*$ selection than the SNR; while variations between $5$ and $20$ dB have minimal impact, detection performance only degrades sharply at very low SNR (e.g., $0$ dB), leading to a shift toward smaller optimal $Q$. }

\begin{figure}[t]
\centering
    \includegraphics[width=1\linewidth]{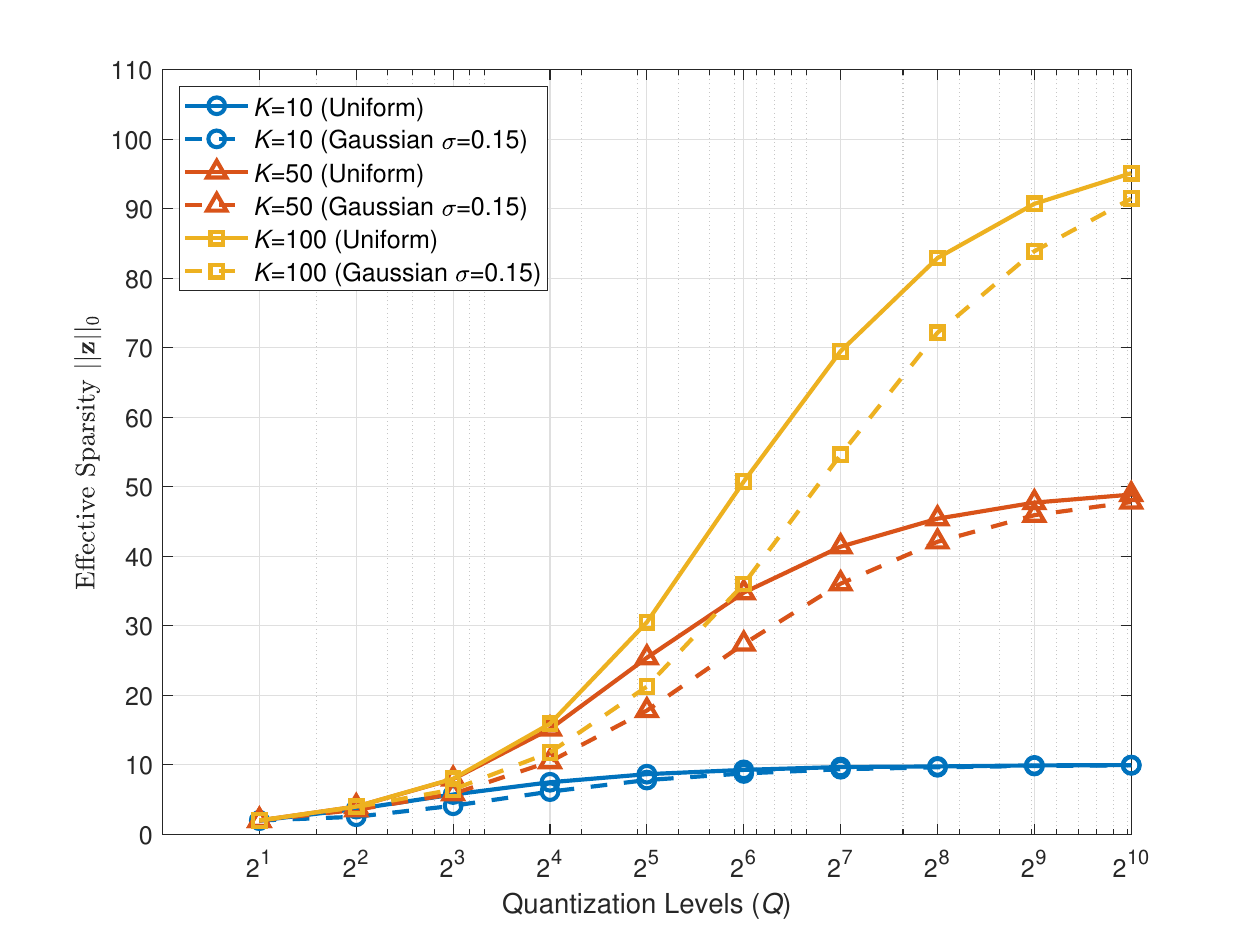}
\caption{Effective sparsity $\|\mathbf{z}\|_0$ versus quantization levels $Q$ for different $K$ and data distributions.}
\label{fig:sparsity}
\vspace{-5mm}
\end{figure}

\subsection{Insight on Sparsity Dynamics}
{Figure \ref{fig:sparsity} explores the dynamics of effective sparsity $\|\mathbf{z}\|_0$ across varying $Q$ and $K$. At low-to-medium resolutions, ``collision-induced sparsification" limits $\|\mathbf{z}\|_0$ significantly below $K$, acting as a self-regulating mechanism that prevents detection complexity from scaling linearly with device density. This ensures system robustness in the quantization-limited region even as $K$ increases. Furthermore, non-uniform data distributions (e.g., truncated Gaussian $\mathcal{N}(0.5, 0.15^2)$) further intensify this effect compared to the uniform case. As $Q$ grows, $\|\mathbf{z}\|_0$ converges to $K$, escalating detection challenges within the high-dimensional search space. This behavior confirms that using $K$ in the MSE bound (\ref{eq:lasso_matrix_bound}) provides a robust, worst-case estimate and justifies the existence of an optimal $Q^*$ that balances resolution fidelity against the challenges of sparse recovery across varying AIoT network densities.}

\begin{figure}[t]
\centering
    \includegraphics[width=1\linewidth]{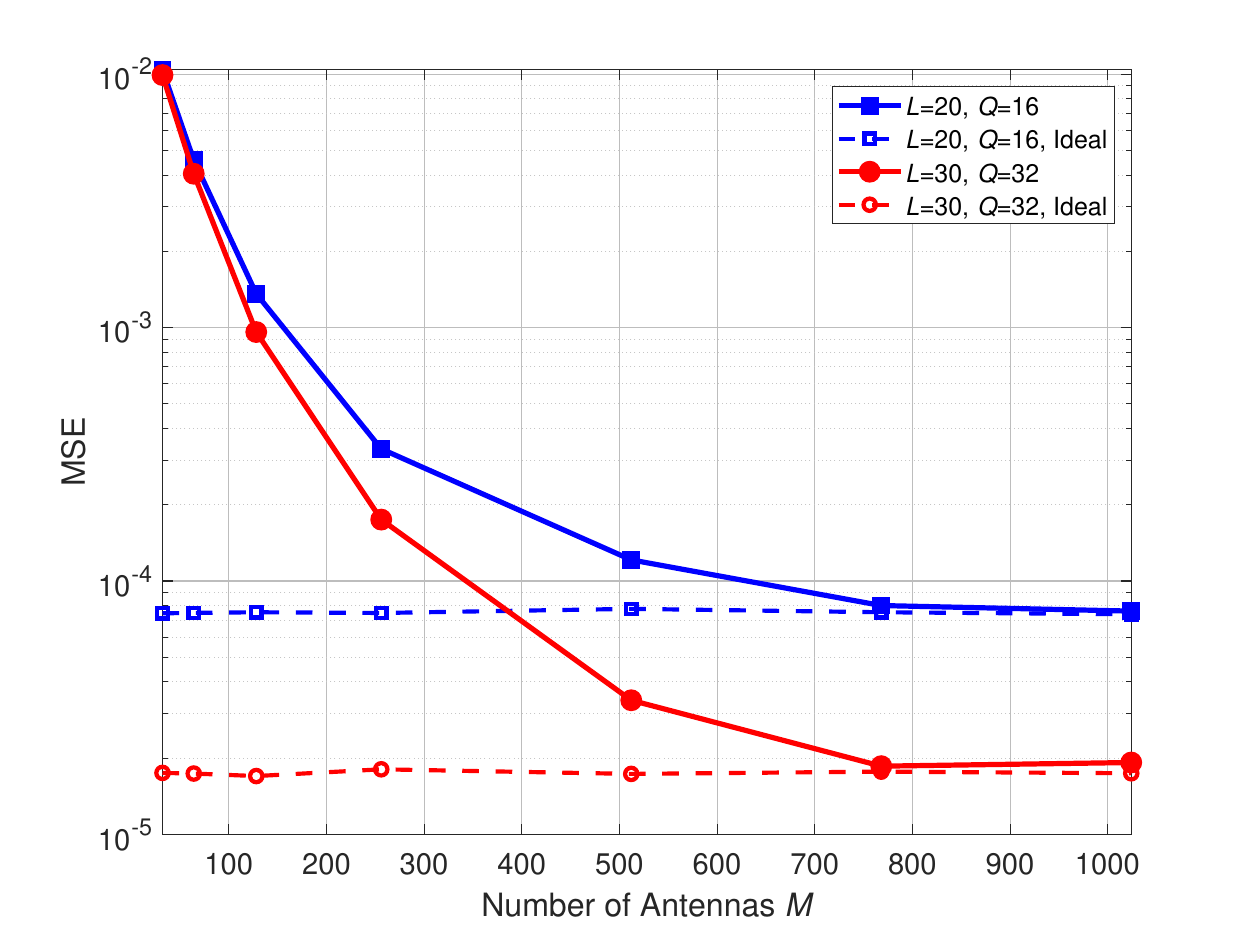}
\caption{MSE versus the number of receive antennas \( M \).}
\label{fig:Mchange}
\vspace{-5mm}
\end{figure}

\subsection{Effect of Antenna Number \( M \)}
{Figure \ref{fig:Mchange} illustrates the MSE performance as a function of the number of receive antennas $M$ at $\text{SNR} = 20$ dB. As $M$ increases, the channel hardening effect becomes more pronounced, leading to a more stable composite channel gain and a corresponding reduction in MSE. Notably, the experimental results in Figure \ref{fig:Mchange} are shown to asymptotically approach the ``Ideal'' baseline—which represents a pure AWGN scenario with constant SNR. This convergence proves that in practical massive MIMO scales, the proposed MD-AirComp+ framework effectively mitigates channel fading and achieves near-optimal computation accuracy. Most importantly, it demonstrates that such high performance can be maintained without the overhead of acquiring precise individual CSI, validating the robustness of our blind mitigation strategy.}

\begin{figure}[t]
\centering
    \includegraphics[width=1\linewidth]{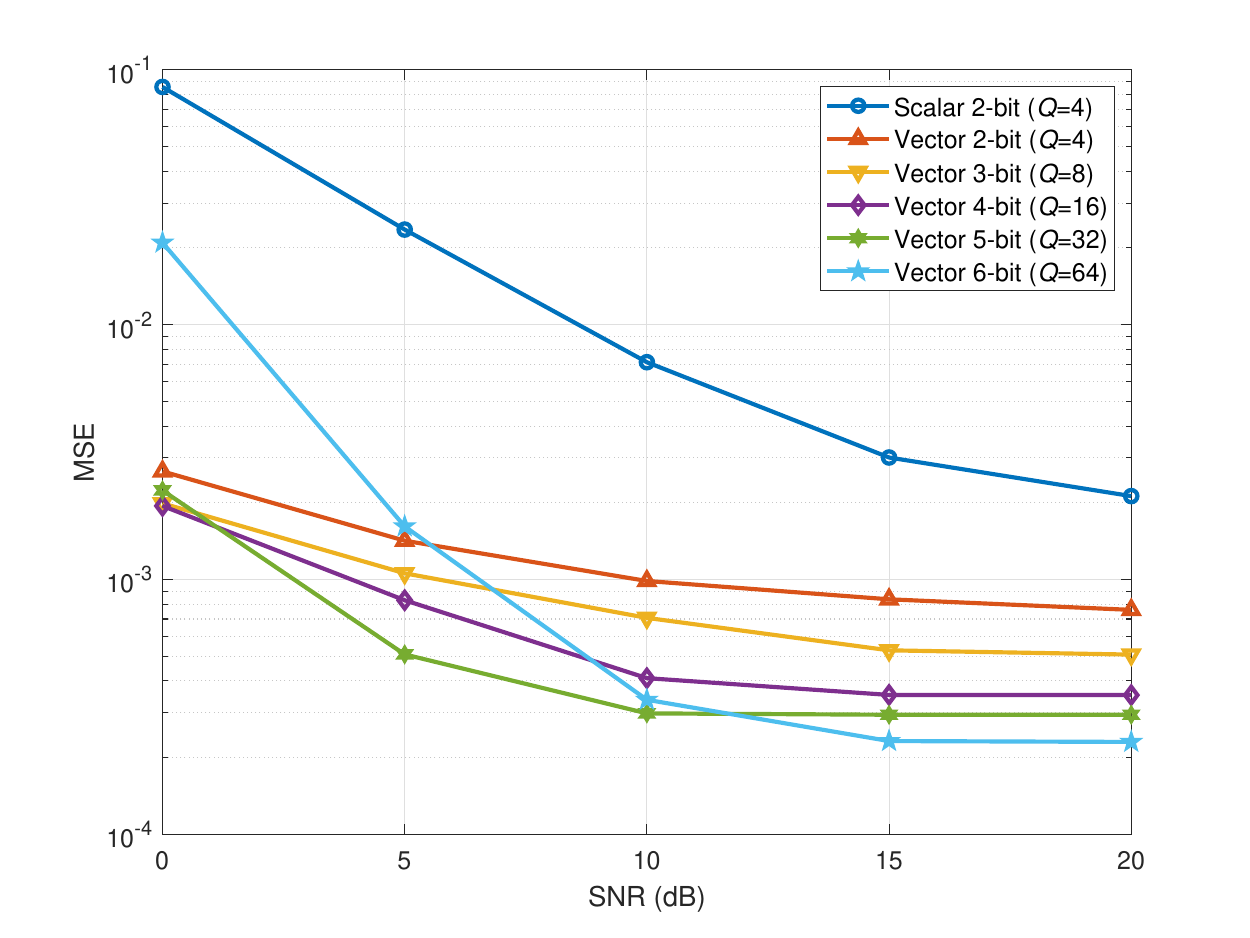}
\caption{Comparison of MSE vs. SNR for various quantization schemes with the same total channel uses.}
\label{fig:Quant}
\vspace{-5mm}
\end{figure}

\subsection{Vector-Quantized MD-AirComp+}
The extension of scalar quantization, as outlined in (\ref{eq:quant}), to VQ is straightforward and does not change the system model (\ref{eq:sysmod}) or the detection framework. However, we hypothesize that VQ-based MD-AirComp+ offers superior computational accuracy compared to the scalar quantization counterpart, particularly in bandwidth-constrained environments.

To substantiate this claim, we simulate a scenario inspired by federated conformal prediction \cite{zhu2024federated}. Each device independently generates a $10$-dimensional probability vector, which corresponds to the softmax output in classification tasks. The simulation setup considers $K=10$ devices and a fixed total of $L=40$ channel uses. In the scalar quantization baseline, the vector is transmitted element-wise, using $\frac{L}{10}=4$ channel uses per element (thus, the scalar quantization level is limited to $Q=4$). {In contrast, for the VQ approach, a shared codebook of $Q$ centroids $\{\mathbf{c}_1, \dots, \mathbf{c}_Q\}$ is pre-constructed via K-means clustering on a representative probability simplex \cite{qiao2024massive}. Each device maps its 10-dimensional vector to the nearest centroid and transmits the index in one shot using the full $L=40$ resources. Since the codebook is shared, the server's task remains the recovery of the frequency vector $\mathbf{z}$ (i.e., the count of devices assigned to each centroid). This demonstrates that our detection framework is inherently agnostic to the dimensionality of the source, effectively leveraging the structured nature of VQ to reduce distortion within the same bandwidth constraint.}

We evaluate multiple VQ codebook sizes, $Q \in \{4, 8, 16, 32, 64\}$, to assess the impact on performance. Depending on the relationship between $L$ and $Q$, the modulation codebook and detection algorithm are chosen accordingly. Specifically, when $L \ge Q$, the $Q$ columns of the modulation codebook $\mathbf{P}$ are selected from an $L \times L$ normalized discrete Fourier transform matrix, and a matched-filter receiver is used. When $L < Q$, a submatrix from ${\bf P_0}$ is used, and signal recovery is carried out via Algorithm~\ref{alg2}. The numerical results are presented in Figure \ref{fig:Quant}.

As illustrated in Figure \ref{fig:Quant}, for a total channel uses of $L=40$, all VQ-based schemes outperform the scalar quantized approach in terms of computing MSE. This performance enhancement stems from the exploitation of the structured nature of the probability vectors in the VQ scheme, allowing for lower distortion, especially at very low rates. Notably, when $Q=64$, the detection problem becomes underdetermined, leading to a degradation in MSE performance at low SNRs, specifically for SNR $<10$ dB. This observation highlights the inherent trade-off between computational accuracy and communication overhead in VQ-based setups, where excessively large codebook sizes may introduce additional detection errors, thus compromising overall performance.

\section{Conclusion}
We propose the MD-AirComp+ framework for efficient uplink computation in large-scale networks, leveraging massive MIMO properties to avoid channel pre-equalization. The framework adaptively selects the optimal quantization level, balancing accuracy and communication overhead. Through MSE analysis, we identify the optimal quantization level that minimizes computing error under limited resources. Additionally, a deep unfolding method is introduced to reduce receiver-side detection complexity, achieving 25x faster convergence with enhanced performance through prior information. Our approach offers an efficient and scalable solution for applications like collaborative inference and federated conformal predictions in wireless networks.

\end{document}